\documentclass[epj]{svjour}
\usepackage{amscd}
\usepackage{amssymb}
\usepackage{amsmath}
\usepackage{dcolumn}
\usepackage[T1]{fontenc}
\usepackage[latin1]{inputenc}
\usepackage{graphicx}
\usepackage{graphics}
\usepackage{color}
\usepackage{latexsym}
\usepackage{amsfonts}
\def\be{\begin{equation}}
\def\ee{\end{equation}}
\def\ba{\begin{eqnarray}}
\def\ea{\end{eqnarray}}
\def\>{\rangle}
\def\<{\langle}
\def\n{\nonumber}

\begin{document}
\title{Quantum Brownian motion and\\the second law of thermodynamics}
\author{ILki Kim\inst{1}\thanks{\emph{e-mail}: hannibal.ikim@gmail.com}
        \and G\"{u}nter Mahler\inst{2}}
\institute{Department of Physics, North Carolina Central University,
           Durham, NC 27707, U.S.A. \and
           Institute of Theoretical Physics I,
           University of Stuttgart, Pfaffenwaldring 57/IV, 70550 Stuttgart, Germany}
\date{\today}
\abstract{We consider a single harmonic oscillator coupled to a bath
at zero temperature. As is well known, the oscillator then has a
higher average energy than that given by its ground state. Here we
show analytically that for a damping model with arbitrarily discrete
distribution of bath modes and damping models with continuous
distributions of bath modes with cut-off frequencies, this excess
energy is less than the work needed to couple the system to the
bath, therefore, the quantum second law is not violated. On the
other hand, the second law may be violated for bath modes without
cut-off frequencies, which are, however, physically unrealistic
models.
}
\maketitle
%
\section{Introduction}
Thermodynamics originally developed as a purely phenomenological
description of the effects caused by changes in temperature,
pressure, and volume on physical systems at the macroscopic scale.
At the heart of thermodynamics there are four well-known laws
\cite{CAL85}; the zeroth law allows us to define temperature scales
and thermometers while the first law is nothing else than a
generalized expression of the law of energy conservation. The second
law introduces the concept of thermodynamic entropy, which never
decreases for an isolated system. The third law states that as a
system approaches the zero temperature, the entropy of the system
approaches zero. Later on, Boltzmann and his followers created and
developed statistical thermodynamics by reducing the
phenomenologically described thermodynamics entirely to the scheme
of classical statistical mechanics. When quantum mechanics appeared,
the statistical thermodynamics had to take into account additional
factors offered by quantum mechanics, but the overall structure of
thermodynamics, its fundamental laws, and its meaning fit for
macroscopic systems remained unchanged since quantum mechanics was
believed to play no roles at the macroscopic scale.

A big challenge for thermodynamics arose with the miniaturization of
a system under consideration \cite{SPI05}; in contrast to common
quantum statistical mechanics which is intrinsically based on a
vanishingly small coupling between system and bath, the finite
coupling strength between them causes some subtleties that must be
recognized. Recent advances in technology have enabled us to
experimentally study mesoscopic systems and test various fundamental
concepts. The field of nano electro-mechanical systems (NEMS)
especially has emerged with a great potential, e.g., in quantum
limit detection and amplification \cite{LAH04,CLE04}, and {\em
welcher-Weg} (`which-path') interferometry \cite{MAC03}. Here, the
effects of dissipative environments that are negligible in
macroscopic resonators become detrimental, and the noise is,
therefore, a major limiting factor in control of NEMS resonators.
Theoretically, NEMS resonators can be modeled as the simplest form
in the scheme of quantum Brownian motion (see \cite{HAE05} for
fundamental aspects of quantum Brownian motion). Such a development
in various fields related to the quantum statistical and mesoscopic
physics has led to considerable interest in the area of quantum and
mesoscopic thermodynamics, especially with the question raised on
the validity of the thermodynamic laws. Discussions about what is
the meaning of quantum thermodynamics \cite{SPI05,MAH04} have
started and continued up to now.

The validity of the second law was questioned in the scheme of
quantum Brownian motion \cite{SPI05}, motivated from the observation
of the fact that a single harmonic oscillator coupled to a bath at
zero temperature has indeed a higher average energy value than the
uncoupled harmonic oscillator ground state (see also \cite{NAG02}),
which could not be in accordance with the second law in its
Kelvin-Planck form \cite{CAL85} that {\em a system operating in
contact with a thermal reservoir cannot produce positive work in its
surroundings} ({\em cf.} for a discussion on the validity of the
quantum third law in the low temperature limit, see, e.g., Refs.
\cite{FOR05}, \cite{HAE06}). However, this argument has been shown
to be wrong by Ford and O'Connell \cite{FOR06}; by means of the
generalized Langevin equation they showed, for the well-known Drude
model for the spectral density of bath modes, that the apparent
excess energy in the coupled harmonic oscillator, however, cannot be
used to extract useful work since the minimum value of the work to
couple the free oscillator to a bath takes above and beyond this
excess energy, therefore, the second law of thermodynamics is
inviolate even in the quantum regime ({\em i.e.}, for cases with
non-negligible coupling strengths at temperature $T = 0$ without
thermal fluctuation). Unfortunately they were unable to explicitly
connect their result with its model-independent, deep quantum
origin, thus the validity of the {\em quantum} second law for a more
general form of the spectral density of bath modes, $J(\omega)$
would still remain an open question; actually, in the experimental
study of mesoscopic systems one might be able to manipulate the
spectral density $J(\omega)$, to some extent, in his own way. In
this paper, we would like to discuss the second law for various
damping models. We will first show the validity of the second law
for a discrete distribution of bath modes by exactly proving the
second-law inequality in a simple form obtained from the general
treatment of the susceptibility (see Sec. \ref{sec:2nd_law}).
Subsequently, the inequality will be appropriately applied for
various continuous distributions of bath modes. It is then found
that for damping models with cut-off frequencies, the second law
holds, whereas interestingly, we may have its violation for damping
models with cut-off frequency-free $J(\omega)$, which are, however,
physically unrealistic (see Sec. \ref{sec:models}). Let us begin
with a brief review on the basics of the quantum Brownian motion. We
will below adopt the notations used in \cite{ING98}.

\section{Basics and its general treatment}\label{sec:basics}
The quantum Brownian motion in consideration is described by the
model Hamiltonian
\begin{equation}\label{eq:total_hamiltonian1}
    \hat{H}\; =\; \hat{H}_s\, +\, \hat{H}_b\, +\,
    \hat{H}_{sb}\,,
\end{equation}
where
\begin{eqnarray}\label{eq:total_hamiltonian2}
    \hat{H}_s &=& \frac{\hat{p}^2}{2 M} +
    \frac{M}{2}\,\omega_0^2\,\hat{q}^2\,;\,
    \hat{H}_b\, =\, \sum_{j=1}^N \left(\frac{\hat{p}_j^2}{2 m_j} +
    \frac{m_j}{2} \omega_j^2\,\hat{x}_j^2\right)\n\\
    \hat{H}_{sb} &=& -\hat{q} \sum_{j=1}^N c_j\,\hat{x}_j\, +\, \hat{q}^2
    \sum_{j=1}^N \frac{c_j^2}{2 m_j\,\omega_j^2}\,.
\end{eqnarray}
Here, from the hermiticity of Hamiltonian, the coupling constants
$c_j$ are obviously real-valued. Without any loss of generality, we
assume that
\begin{equation}\label{eq:frequency_relation1}
    \omega_1\, \leq\, \omega_2\, \leq\, \cdots\, \leq\, \omega_{N-1}\, \leq\, \omega_N\,.
\end{equation}
By means of the Heisenberg equation of motion for $\hat{p}$ we can
derive the quantum Langevin equation
\begin{equation}\label{eq:eq_of_motion1}
    \textstyle M\,\ddot{\hat{q}} \, +\,
    M \int_0^t d s\, \gamma(t -s)\, \dot{\hat{q}}(s)\,
    +\, M\,\omega_0^2\,\hat{q}\; =\; \hat{\xi}(t)\,,
\end{equation}
where we used $\hat{p} = M \dot{\hat{q}}$, and the damping kernel
and the noise operator are respectively given by
\begin{eqnarray}\label{eq:damping_kernel1}
    \displaystyle \gamma(t) &=& \frac{1}{M} \sum_{j=1}^N
    \frac{c_j^2}{m_j\,\omega_j^2}\cos(\omega_j\,t)\,;\,
    \displaystyle \hat{\xi}(t)\, =\, - M \gamma(t)\,\hat{q}(0) +\n\\
    && \sum_{j=1}^N c_j \left\{\hat{x}_j(0) \cos(\omega_j\,t)\, +\,
    \frac{\hat{p}_j(0)}{m_j\,\omega_j} \sin(\omega_j\,t)\right\}\,.
\end{eqnarray}
Introducing the spectral density of bath modes as a characteristic
of the bath,
\begin{equation}\label{eq:spectral_density1}
    J(\omega)\; =\; \pi \sum_{j=1}^N \frac{c_j^2}{2 m_j\,
    \omega_j}\,\delta(\omega - \omega_j)\,,
\end{equation}
we can express the damping kernel as
\begin{equation}\label{eq:damping_kernel2}
    \gamma(t)\; =\; \frac{2}{M} \int_0^{\infty} \frac{d \omega}{\pi}
    \frac{J(\omega)}{\omega} \cos(\omega\,t)\,.
\end{equation}

Let us apply the Laplace transform to eq. (\ref{eq:eq_of_motion1})
with the aid of \cite{IKI06,ROB66}
\begin{eqnarray}
    &&\mathcal{L}\{\cos(\omega_j\,t)\}(s)\; =\; {\frac{s}{s^2 +
    \omega_j^2}}\,,\label{eq:laplace_transform_cos_sin}\\
    &&\mathcal{L}\{\sin(\omega_j\,t)\}(s)\; =\; {\frac{\omega_j}{s^2 +
    \omega_j^2}}\,.
\end{eqnarray}
With $s = -i \omega + 0^+ = -i (\omega + i\,0^+)$ we then easily
obtain
\begin{eqnarray}\label{eq:laplace_q_final}
    \hspace*{-.5cm}\hat{q}_{\omega} &:=& \mathcal{L}\{\hat{q}(t)\}(-i \omega + 0^+)\n\\
    \hspace*{-.5cm} &=& \tilde{\chi}(\omega)\,\left[\hat{\xi}_{\omega} - i \omega
    M \{1 + \tilde{\gamma}(\omega)\}\,\hat{q}(0) + M \dot{\hat{q}}(0)\right]\,,
\end{eqnarray}
where the Laplace-transformed damping kernel, the dynamic
susceptibility, and the Laplace-transformed noise operator are,
respectively, given by
\begin{eqnarray}
    \hspace*{-.5cm}&&\tilde{\gamma}(\omega)\, =\, \frac{i \omega}{M} \sum_j^N \frac{c_j^2}{m_j\,\omega_j^2}\,
    \frac{1}{\omega^2 - \omega_j^2}\,,\label{eq:gamma_tilde1}\\
    \hspace*{-.5cm}&&\tilde{\chi}(\omega)\, =\, \frac{1}{M}\,\frac{1}{\omega_0^2 -\omega^2 -
    i \omega\,\tilde{\gamma}(\omega)}\,,\label{eq:gamma_tilde1_2}\\
    \hspace*{-.5cm}&&\hat{\xi}_{\omega}\, =\, \sum_{j=1}^{N} \frac{c_j}{\omega^2 - \omega_j^2} \left\{i \omega\,\hat{x}_j(0)
    - \frac{\hat{p}_j(0)}{m_j}\right\} - M \tilde{\gamma}(\omega)\,\hat{q}(0)\,.\n
\end{eqnarray}
Substituting (\ref{eq:gamma_tilde1}) into (\ref{eq:gamma_tilde1_2}),
we get
\begin{equation}\label{eq:susceptibility1}
    \tilde{\chi}(\omega)\; =\; \frac{\displaystyle -\frac{1}{M}
    \prod_{j=1}^N\, (\omega^2 - \omega_j^2)}{D_{\tilde{\chi}}(\omega)}\,,
\end{equation}
where
\begin{equation}\label{eq:susceptibility_denominator1}
    \hspace*{-.1cm}D_{\tilde{\chi}}(\omega)\, =\, {\displaystyle
    \prod_{j=0}^N (\omega^2 - \omega_j^2) - \frac{\omega^2}{M}
    \sum_{j=1}^N \frac{c_j^2}{m_j\,\omega_j^2} \prod_{j'=1 \atop (\ne j)}^N
    \left(\omega^2 - \omega_{j'}^2\right)}\,.
\end{equation}
It is known \cite{LEV88} that the susceptibility
$\tilde{\chi}(\omega)$ in (\ref{eq:gamma_tilde1_2}) has poles at the
normal-mode frequencies of the total system $\hat{H}$ in
(\ref{eq:total_hamiltonian1}), $\pm\,\bar{\omega}_k$ with $k = 0, 1,
2, \cdots, N$, so that
\begin{equation}\label{eq:normal_mode_frequency1}
    \omega_0^2\, -\, \bar{\omega}_k^2\, -\, i\,\bar{\omega}_k\,\tilde{\gamma}(\bar{\omega}_k)\; =\;
    0\,.
\end{equation}
Here, we might be able to say that a specific $k = k_0$ would
represent the ``system harmonic oscillator'' with the normal-mode
frequency $\bar{\omega}_{k_0}$, {\em uncoupled} to the ``bath''
consisting of the remaining oscillators with $\bar{\omega}_k$, where
$k \ne k_0$.
From eqs. (\ref{eq:gamma_tilde1_2}), (\ref{eq:susceptibility1}), and
(\ref{eq:normal_mode_frequency1}), we have a compact expression of
the susceptibility,
\begin{equation}\label{eq:susceptibility2}
    \tilde{\chi}(\omega)\; =\; -\frac{1}{M}\, \frac{\displaystyle
    \prod_{j=1}^N\, (\omega^2 - \omega_j^2)}{\displaystyle \prod_{k=0}^N\, (\omega^2 -
    \bar{\omega}_k^2)}\,.
\end{equation}
Without any loss of generality, we here assume that
\begin{equation}\label{eq:frequency_relation2}
    \bar{\omega}_0\, \leq\, \bar{\omega}_1\, \leq\, \cdots\, \leq\,
    \bar{\omega}_{N-1}\, \leq\, \bar{\omega}_N\,.
\end{equation}

The damping function $\tilde{\gamma}(\omega)$ in the frequency
domain has, besides eq. (\ref{eq:gamma_tilde1}), another expression
which is suitable for the case of a continuous distribution of bath
modes; from eqs. (\ref{eq:damping_kernel2}) and
(\ref{eq:laplace_transform_cos_sin}) we obtain
\begin{eqnarray}
    \hspace{-.5cm}&&{\textstyle \tilde{\gamma}(\omega)\; =\; \frac{i}{M} \int_0^{\infty} \frac{d \omega'}{\pi}
    \frac{J(\omega')}{\omega'} \left(\frac{1}{\omega' + \omega}\, -\, \frac{1}{\omega' -
    \omega}\right)\,,}\label{eq:damping_kernel2_1}\\
    \hspace{-.5cm}&&{\textstyle \tilde{\gamma}(\omega)\big{|}\vspace*{2.5cm}_{\omega \to \atop \omega +
    i\,0^{+}}\; =\; \frac{J(\omega)}{M\, \omega}\, +}\n\\
    \hspace{-.5cm}&&{\textstyle \hspace*{1.3cm}\frac{i}{M} \int_0^{\infty} \frac{d \omega'}{\pi}
    \frac{J(\omega')}{\omega'}\; P\left(\frac{1}{\omega' + \omega} - \frac{1}{\omega' -
    \omega}\right)\,.}\label{eq:damping_kernel3}
\end{eqnarray}
We here used the well-known formula $1/(x + i\,0^+) = P(1/x) - i \pi
\delta(x)$ for $x = \omega' - \omega$. For the simple Ohmic case
$J_{0}(\omega)\, =\, M \gamma_o\,\omega$ with an
$\omega$-independent constant $\gamma_o$, we easily have
$\gamma_0(t) = 2 \gamma_o\,\delta(t)$, and
$\tilde{\gamma}_{0}(\omega) = \gamma_o$ with a vanishing principal
({\em or} imaginary) part in (\ref{eq:damping_kernel3}), while for
the Drude model where $J_d(\omega)\, =\,
M\,\gamma_o\,\omega\,\omega_d^2/(\omega^2 + \omega_d^2)$ with a
cut-off frequency $\omega_d$, we have $\gamma_d(t) = \gamma_o\,
\omega_d\, e^{-\omega_d\,t}$, and
\begin{equation}\label{eq:drude_gamma1}
    \tilde{\gamma}_d(\omega)\; =\; \frac{\gamma_o\, \omega_d^2}{\omega^2 + \omega_d^2}\; +\;
    i\,\frac{\gamma_o\, \omega_d\, \omega}{\omega^2 + \omega_d^2} \; =\;
    \frac{\gamma_o\, \omega_d}{\omega_d - i \omega}\,.
\end{equation}

For a later purpose, it is interesting to compare
$D_{\tilde{\chi}}(\omega)$ in eq.
(\ref{eq:susceptibility_denominator1}) with the denominator of the
right hand side in (\ref{eq:susceptibility2}). Then, we can easily
find that
\begin{equation}\label{eq:relation_among_normal_modes1}
    \sum_{k=0}^N \bar{\omega}_k^2\; =\; \sum_{j=0}^N
    \omega_j^2\, +\, \gamma(0)\,;\; \prod_{k=0}^N \bar{\omega}_k^2\; =\; \prod_{j=0}^N
    \omega_j^2\,.
\end{equation}
Here, $\gamma(0) = \gamma(t)|_{t=0} \geq 0$ in eq.
(\ref{eq:damping_kernel1}). From this comparison of the denominators
at $\omega = \omega_N$, we also obtain $D_{\tilde{\chi}}(\omega_N)
\leq 0$ and so $\omega_N \leq \bar{\omega}_N$. Similarly, we can
acquire both $D_{\tilde{\chi}}(\omega_1) \leq 0$ for $N$ odd and
$D_{\tilde{\chi}}(\omega_1) \geq 0$ for $N$ even, which lead to the
fact that $\bar{\omega}_0 \leq \omega_1$ for any given $N$. Further,
we can obtain the relationship, $D_{\tilde{\chi}}(\omega_j) \cdot
D_{\tilde{\chi}}(\omega_{j-1}) \leq 0$ for any $j$. Therefore, it is
found that
\begin{equation}\label{eq:frequency_relation3}
    \bar{\omega}_0\, \leq\, \omega_1\, \leq\, \bar{\omega}_1\, \leq\, \cdots\,
    \leq\, \omega_{N-1}\, \leq\, \bar{\omega}_{N-1}\, \leq\, \omega_N\, \leq\,
    \bar{\omega}_N\,.
\end{equation}
By using $D_{\tilde{\chi}}(\omega_0)$, we can also show that
$\bar{\omega}_0 \leq \omega_0 \leq \bar{\omega}_N$ (see also
\cite{FOR88}). Within this {\em general} treatment of the
susceptibility, we would like to consider the quantum second law
below.

\section{General validity of the quantum second law (discrete bath modes)}\label{sec:2nd_law}
The energy of the system oscillator $\hat{H}_s$ at zero temperature
can be calculated by means of the partition function $Z =
\text{Tr}\,e^{-\beta \hat{H}}$ with $\beta = 1/k_B T$ as
\begin{equation}\label{eq:excess_energy0}
    {\textstyle \left\<\hat{H}_s\right\>_{T=0}\; =\;
    \left.\frac{\text{Tr}\left(\hat{H}_s\,
    e^{-\beta \hat{H}}\right)}{Z}\right|_{\beta \to \infty}\; =:\;
    E_s(0)\,.}
\end{equation}
It is well-known \cite{MAH04,WEI99} that the system-bath
entanglement induced by the coupling term $\hat{H}_{sb}$ in
(\ref{eq:total_hamiltonian1}) leads to the fact that the system
oscillator $\hat{H}_s$, initially in a pure state (here, its ground
state with the minimum energy $E_g = \hbar\,\omega_0/2$), is not in
the pure state any longer but in a mixed state with a fluctuation in
energy, and so we actually have $E_s(0) > E_g$. It was even
discussed in \cite{BUT05} that the energy fluctuation measurements
can provide entanglement information ({\em cf}. for a
thermodynamical approach to quantifying entanglement in bipartite
qubit states, see \cite{OPP02}). From the fluctuation-dissipation
theorem \cite{ING98}, we can also easily obtain
\begin{equation}\label{eq:excess_energy1}
    E_s(0)\; =\; \frac{M \hbar}{2 \pi} \int_0^{\infty} d\omega\,(\omega_0^2\,
    +\, \omega^2)\; \text{Im}\,\tilde{\chi}(\omega + i\,0^+)\,.
\end{equation}
The factor $\text{Im}\,\tilde{\chi}(\omega + i\,0^+)$ can be
evaluated from eq. (\ref{eq:susceptibility2}) with $\omega \to
\omega + i\,0^+$. By means of the technique used, e.g., in
\cite{KAM04}, eq. (\ref{eq:excess_energy1}) can be rewritten as
\begin{equation}\label{eq:contour_integral1}
    E_s(0)\; =\;
    \frac{\hbar}{2}\, \frac{1}{2 \pi i} \oint d\omega\, \frac{\omega_0^2\,
    +\, \omega^2}{G(\omega)}\,,
\end{equation}
where $G(\omega) = - 1/\{M\,\tilde{\chi}(\omega)\}$, and the
integration path is a loop around the {\em positive real axis} in
the complex $\omega$-plane, consisting of the two branches, $(\infty
+ i \epsilon,\,i \epsilon)$ and $(-i \epsilon,\,\infty - i
\epsilon)$. Therefore, $E_s(0)$ can be exactly obtained in closed
form from the residues evaluated at all zeroes of $G(\omega)$ on the
positive real axis. It is also interesting to note that the
entanglement between any pair of the bath oscillators $\hat{H}_j =
\hat{p}_j^2/2\,m_j + m_j\,\omega_j^2\,\hat{x}_j^2/2$ with $j= 1, 2,
3, \cdots, N$ is induced by the system-bath entanglement and the
well-known entanglement swapping.\cite{ALB01} As a result, we must
obtain an excess energy for any $j$, i.e., $\<\hat{H}_j\>_{T=0} >
\hbar\,\omega_j/2$. However, the energy of the total system,
$\<\hat{H}\>_{T=0} = \sum_{k = 0}^N\, \hbar\, \bar{\omega}_k/2$ is
clearly not equivalent to $\<\hat{H}_s\>_{T=0}\, +\, \sum_{j=1}^N\,
\<\hat{H}_j\>_{T=0} = \<\hat{H}_s\, +\, \hat{H}_b\>_{T=0}$.

The minimum work required to couple a harmonic oscillator at
temperature $T$ to a bath at the same temperature is equivalent to
the Helmholtz free energy of the {\em coupled} total system minus
the free energy of the {\em uncoupled} bath \cite{CAL85,FOR06}. The
Helmholtz free energy can be obtained from the canonical partition
function $Z_s(\beta) = \mbox{Tr}\, e^{-\beta \hat{H}}/\mbox{Tr}_b\,
e^{-\beta \hat{H}_b}$ as $F(T) = -k_B\,T\,\ln Z_s$, where
$\mbox{Tr}_b$ denotes the partial trace for the bath alone (in the
absence of a coupling between system and bath, this would exactly
correspond to the partition function of the system only). By means
of the normal-mode frequencies $\bar{\omega}_k$ the partition
function can be rewritten as
\begin{equation}\label{eq:partion_function1}
    Z_s(\beta)\; =\; \frac{\displaystyle
    \prod_{k=0}\, \sum_{n_k=0}\, e^{-\beta \hbar \bar{\omega}_k
    \left(n_k + \frac{1}{2}\right)}}{\displaystyle \prod_{j=1}\, \sum_{n_j=0}\, e^{-\beta \hbar
    \omega_j \left(n_j + \frac{1}{2}\right)}}
\end{equation}
so that we can easily get, for $\beta \to \infty$,
\begin{equation}\label{eq:helmholtz_energy0}
    F(0)\; =\; \frac{\hbar}{2}\, \left(\sum_{k=0}^N \bar{\omega}_k\, -\,
    \sum_{j=1}^N \omega_j\right)\,.
\end{equation}
With the aid of eq. (\ref{eq:relation_among_normal_modes1}), it is
evidently found that $F(0) > \hbar\,\omega_j/2$ for any $j = 0, 1,
2, \cdots, N$. Further, we have, from \cite{FOR85},
\begin{equation}
    \hspace*{-.3cm}F(T)\; =\; \frac{1}{\pi} \int_0^{\infty}\, d\omega\,
    f(\omega,T)\; \text{Im}\left\{\frac{d}{d \omega} \ln
    \chi(\omega + i 0^+)\right\}\,,\label{eq:helmholtz_energy1}\\
\end{equation}
where $f(\omega,T) = k_B\,T\,\ln \{2\,\sinh (\hbar\,\omega/2\,k_B
T)\}$. Similarly to eq. (\ref{eq:contour_integral1}), we can obtain
an integral form of the free energy at $T = 0$,
\begin{equation}
    F(0)\; =\; \frac{\hbar}{2}\, \frac{1}{2 \pi i} \oint d\omega\,
    \frac{\omega\, G'(\omega)}{G(\omega)}\,.\label{eq:contour_integral2}
\end{equation}
Here, $f(\omega,0) = \hbar\,\omega/2$.

Now, we are in a position to exactly formulate the quantum second
law within this general treatment; from eqs.
(\ref{eq:contour_integral1}) and (\ref{eq:contour_integral2}) with
(\ref{eq:gamma_tilde1_2}), we easily find an expression
\begin{equation}\label{eq:second_law_in_general_treatment1}
    K :=\; F(0) - E_s(0)\; =\; \frac{\hbar}{4 \pi}\, \oint d\omega\,
    \frac{\omega^2\, \tilde{\gamma}'(\omega)}{G(\omega)}\,,
\end{equation}
and, for the validity of the second law, we have to get $K \geq 0$
for any $N$ (the number of the bath oscillators) and the limit $N
\to \infty$. Here, $K$ can exactly be evaluated from all residues of
the integrand on the positive {\em real} axis. Substituting
(\ref{eq:susceptibility2}) with $\tilde{\chi}(\omega) = -1/M
G(\omega)$ and (\ref{eq:gamma_tilde1}) into
(\ref{eq:second_law_in_general_treatment1}), we obtain, after a
fairly lengthy evaluation of the contour integration (see
Appendix~\ref{sec:appendix1} for details), the exact result
\begin{equation}\label{eq:second_law_in_general_treatment2}
    K\; =\; \frac{\hbar}{8 M}\, \sum_{k=0}^N\, \mathcal{A}_k\,,
\end{equation}
where
\begin{eqnarray}\label{eq:second_law_in_general_treatment21}
    \mathcal{A}_k &=& \bar{\omega}_k\,
    \frac{\displaystyle \prod_{j=1}^N\, (\bar{\omega}_k^2 - \omega_j^2)}{\displaystyle
    \prod_{k'=0 \atop (\ne k)}^N\, (\bar{\omega}_k^2 - \bar{\omega}_{k'}^2)}\; \times\\
    && \sum_{l=1}^N\, \frac{c_l^2}{m_l\,\omega_l^2}\;\,
    P\left\{\frac{1}{(\omega_l + \bar{\omega}_k)^2}\, +\, \frac{1}{(\omega_l -
    \bar{\omega}_k)^2}\right\}\,.\n
\end{eqnarray}
Considering each summand $\mathcal{A}_k$ from $k = N$ with keeping
in mind the frequency relationship in
(\ref{eq:frequency_relation3}), we see that each of the summand is
non-negative and so $K \geq 0$ indeed! Separately from this result
for discrete bath modes, we will next discuss the second law for
continuous bath modes. For doing this job, we will consider a
continuation of the spectral density $J(\omega)$ from its original
form in (\ref{eq:spectral_density1}).

\section{The second law for continuous bath modes}\label{sec:models}
For a discussion of the second law for a continuous distribution of
bath modes, we rewrite eq.
(\ref{eq:second_law_in_general_treatment1}) as
\begin{equation}\label{eq:conti1}
    K\; =\; \frac{\hbar}{4 \pi}\, \left(\int_0^{\infty}
    d\omega\, \frac{\omega^2\;
    \tilde{\gamma}'_{-}(\omega)}{G_{-}(\omega)}\,
    -\, \int_0^{\infty} d\omega\,
    \frac{\omega^2\; \tilde{\gamma}'_{+}(\omega)}{G_{+}(\omega)}\right)\,,
\end{equation}
where the subscripts $+/-$ denote the branches $(\infty + i
\epsilon,\,i \epsilon)$ and $(-i \epsilon,\,\infty - i \epsilon)$,
respectively, so that
\begin{eqnarray}\label{eq:branch1_1}
    G_{+}(\omega) &:=& {\textstyle G(\omega)\big{|}\vspace*{2.5cm}_{\omega \to \atop \omega +
    i\,0^{+}}\; =\; \omega^2\, -\, \omega_0^2\, +\, i\,\frac{J(\omega)}{M}\; -}\n\\
    && {\textstyle \frac{\omega}{M} \int_0^{\infty} \frac{d \omega'}{\pi}
    \frac{J(\omega')}{\omega'}\; P\left(\frac{1}{\omega' + \omega} - \frac{1}{\omega' -
    \omega}\right)\,,}
\end{eqnarray}
and $G_{-}(\omega) := G(\omega)\big{|}_{\omega \to \atop \omega -
i\,0^{+}} = G_{+}^{\ast}(\omega)$. Here, we used $G(\omega) =
\omega^2\,-\,\omega_0^2\,+\,i\,\omega\,\tilde{\gamma}(\omega)$ with
eq. (\ref{eq:damping_kernel3}) for $\tilde{\gamma}_+(\omega)$.
Therefore, eq. (\ref{eq:conti1}) easily reduces to
\begin{equation}\label{eq:second_law_inequality_conti}
    K\; =\; \frac{\hbar}{2 \pi}\; \mbox{Im} \int_0^{\infty}
    d\omega\, \frac{\omega^2\, R_+'(\omega)}{G_+(\omega)}\,,
\end{equation}
where $R_+(\omega) = -i\,\tilde{\gamma}_+(\omega)$.

First, for the Ohmic case, $(\tilde{\gamma}_+)_{0}(\omega) =
\gamma_o$, which is the prototype for damping, we easily obtain
$K_{0} = 0$. In fact, both $(E_s)_{0}(0)$ and $F_{0}(0)$ have the
logarithmic divergence, however, the same value, namely,
\begin{equation}\label{eq:ohmic_energy_free_energy}
    \textstyle (E_s)_{0}(0)\; =\; F_{0}(0)\; =\;
    \frac{\hbar\,\gamma_o}{2\,\pi}\,\int_0^{\infty} d\omega\,
    \frac{\omega\,(\omega^2\,+\,\omega_0^2)}{(\omega^2\,-\,\omega_0^2)^2\,
    +\, (\gamma_o\,\omega)^2}
\end{equation}
(see also the discussion in the last paragraphs of Secs.
\ref{sec:drude_model} and \ref{sec:exponential_model}). However, the
Ohmic model is not so realistic in its strict form because the
spectral density of bath modes, $J_{0}(\omega) = M \gamma_o\,\omega$
diverges for large frequencies. We therefore introduce a cut-off
frequency $\omega_c$ which leads to $J_c(\omega)$ decaying smoothly
to zero for large frequencies $\omega > \omega_c$. We will first
consider the Drude model, where $J_d(\omega)$ is polynomially
decaying for $\omega > \omega_c = \omega_d$, and next a damping
model with $J_e(\omega)$ being exponentially decaying for $\omega >
\omega_c = \omega_e$. For these damping models, we will be able to
show that $K > 0$. Subsequently, we will also consider two different
damping models without cut-off frequencies $\omega_c$; first, the
extended Ohmic models where the spectral densities $J(\omega)$
diverge polynomially faster than $J_{0}(\omega)$, and secondly, the
extended Drude models with $J_{d,n}(\omega)$ diverging faster or
more slowly than $J_{0}(\omega)$. Interestingly, we will observe $K
< 0$ for some of the cut-off frequency-free damping models (see
Secs. \ref{sec:extended_ohmic_model} and
\ref{sec:extended_drude_model}).

\subsection{Drude model $(d)$}\label{sec:drude_model}
We briefly review the second law in the Drude model considered in
\cite{FOR06}; it is convenient to adopt, in place of $(\omega_0,
\omega_d, \gamma_o)$, the parameters $({\mathbf w}_0, \Omega,
\gamma)$ through the relations
\begin{eqnarray}\label{eq:parameter_change0}
    &\textstyle \omega_0^2\; :=\; {\mathbf w}_0^2\; \frac{\Omega}{\Omega\, +\, \gamma}\,;\;
    \omega_d\; :=\; \Omega\, +\, \gamma\,;&\n\\
    &\textstyle \gamma_o\; :=\; \gamma\, \frac{\Omega\, (\Omega\, +\, \gamma)\,
    +\, {\mathbf w}_0^2}{(\Omega\, +\, \gamma)^2}\,.&
\end{eqnarray}
Substituting eq. (\ref{eq:drude_gamma1}) with
(\ref{eq:parameter_change0}) into eq. (\ref{eq:gamma_tilde1_2}), we
obtain the susceptibility
\begin{eqnarray}
    \hspace*{-.7cm}&&\tilde{\chi}_d(\omega)\n\\
    \hspace*{-.7cm}&=& -\frac{1}{M}\,
    \frac{\omega\, +\, i\,\omega_d}{\omega^3\, +\, i\,\omega_d\, \omega^2\,
    -\, (\omega_0^2\, +\, \gamma_o\, \omega_d)\, \omega\, -\, i\,\omega_0^2\,
    \omega_d}\label{eq:susceptibility_drude01}\\
    \hspace*{-.7cm}&=& -\frac{1}{M}\,
    \frac{\omega\, +\, i\,(\Omega\, +\, z_1\, +\, z_2)}{(\omega\, +\, i \Omega)
    (\omega\, +\, i z_1) (\omega\, +\, i z_2)}\,,\label{eq:susceptibility_drude1}
\end{eqnarray}
where $z_1 = \gamma/2 + i {\mathbf w}_1$ and $z_2 = \gamma/2 - i
{\mathbf w}_1$ with ${\mathbf w}_1^2 = {\mathbf w}_0^2 -
(\gamma/2)^2$. This gives us $(G_+)_d(\omega) =
-1/\{M\,\tilde{\chi}_d(\omega)\}$ for eq.
(\ref{eq:second_law_inequality_conti}). By means of eq.
(\ref{eq:susceptibility_drude1}), we can even obtain the closed
expressions for both $(E_s)_d(0)$ from (\ref{eq:excess_energy1}) and
$F_d(0)$ from (\ref{eq:helmholtz_energy1}). We give the detailed
derivation of these expressions in Appendix~\ref{sec:appendix2},
which will also be used in Sec. \ref{sec:extended_drude_model}. It
has been numerically shown in \cite{FOR06} that $(E_s)_d(0)$ in
(\ref{eq:energy_drude1}) is actually greater than $E_g =
\frac{\hbar\,{\mathbf w}_0}{2}\sqrt{\frac{\Omega}{\Omega\, +\,
\gamma}}$, and $F_d(0)$ in (\ref{eq:helmholtz_energy2_appendix}) is
even greater than the $(E_s)_d(0)$, i.e., $K_d > 0$. For a later
purpose, we will also evaluate $K_d$ explicitly for various pairs
$(\omega_0, \omega_d)$ (see Table \ref{tab:table2} in Sec.
\ref{sec:extended_drude_model}).

It is noted that in the limit $\omega_d \to \infty$ (equivalently,
$\Omega \to \infty$), we have $K_d \to \frac{\gamma}{\pi {\mathbf
w}_0} E_g$ (see Appendix~\ref{sec:appendix2}). From the comparison
between $\tilde{\gamma}_d(\omega)$ and $\tilde{\gamma}_{0}(\omega)$
({\em or}, equivalently, $J_d(\omega)$ and $J_0(\omega)$), this
result would be interpreted as $K_{0} \to \frac{\gamma}{\pi {\mathbf
w}_0} E_g$. However, it is misleading; $\tilde{\gamma}_d(\omega)$
behaves only for small frequencies, $\omega \ll \omega_d$, like in
the Ohmic case, which corresponds to $\gamma_d(t) \to \gamma_{0}(t)$
only for large times. Actually, $\gamma_d(t) = \gamma_o\,\omega_d\,
e^{-\omega_d\,t}$ with $\omega_d \to \infty$ does not reduce to
$\gamma_{0}(t) = 2\,\gamma_o\,\delta(t) = \lim_{\omega_d \to \infty}
\frac{2}{\sqrt{\pi}}\,\gamma_o\,\sqrt{\omega_d}\,
e^{-\omega_d\,t^2}$. For the evaluation of $K$, however, all
frequencies, $0 \leq \omega < \infty$, have to be considered.
Therefore, we evidently get $\lim_{\omega_d \to \infty}\,K_d \gneq
K_{0} = 0$ .

\subsection{Exponentially decaying model $(e)$}\label{sec:exponential_model}
We now consider a damping model with $J_e(\omega) =
M\,\gamma_o\,\omega\, e^{-\omega/\omega_e}$ which, in the limit
$\omega_e \to \infty$, clearly reduces to $J_{0}(\omega)$ for small
frequencies. Substituting this into eq. (\ref{eq:damping_kernel2}),
we can obtain
\begin{equation}\label{eq:exponential_decay1}
    \gamma_e(t)\; =\; \frac{2}{\pi}\, \frac{\gamma_o\,
    \omega_e}{1\,+\,(\omega_e\,t)^2}\,.
\end{equation}
Applying the Laplace transform \cite{ROB66} to eq.
(\ref{eq:exponential_decay1}) with $s = -i\,\omega + 0^{+}$, it can
be found that
\begin{eqnarray}\label{eq:exponential_frequency1}
    \tilde{\gamma}_e(\omega) &=& \gamma_o\, e^{-\omega/\omega_e}\;
    +\\
    && i\, \frac{\gamma_o}{\pi} \left\{e^{\omega/\omega_e}\,
    E_1\left(\frac{\omega}{\omega_e}\right)\, +\,
    e^{-\omega/\omega_e}\,
    \mbox{Ei}\left(\frac{\omega}{\omega_e}\right)\right\}\,,\n
\end{eqnarray}
(see Appendix \ref{sec:appendix3} for the detailed derivation). By
using this with $E_1'(y) = -E_0(y) = -e^{-y}/y$, we can easily get
$(R_{+}')_e(\omega)$ and $(G_{+})_e(\omega)$, and then introducing a
dimensionless variable $\lambda = \omega/\omega_e$, we arrive at the
expression
\begin{equation}\label{eq:exponential_K_q}
    K_e\; =\; \frac{\hbar\,\gamma_o\,\omega_e^2}{2\,\pi^2}\;\, \mbox{Im}
    \int_0^{\infty} d\lambda\ \frac{f_1(\lambda)}{f_2(\lambda)}\;,
\end{equation}
where
\begin{eqnarray}
    f_1(\lambda) &=& \lambda^2\, \{e^{\lambda}\, E_1(\lambda)\, -\,
    e^{-\lambda}\, \mbox{Ei}(\lambda)\, +\, i\,\pi e^{-\lambda}\}\,,\\
    f_2(\lambda) &=& \omega_e^2\,\lambda^2\, -\, \omega_0^2\, -\,
    \frac{\gamma_o\,\omega_e}{\pi}\, \lambda\, \{e^{\lambda}\, E_1(\lambda)\,
    +\, e^{-\lambda}\, \mbox{Ei}(\lambda)\}\n\\
    && +\, i\,\omega_e\,\gamma_o\,\lambda\,e^{-\lambda}\,.
\end{eqnarray}
We numerically evaluate the integration in
(\ref{eq:exponential_K_q}) for various pairs $(\gamma_o, \omega_e)$
to show that $K_e > 0$ (see Table~\ref{tab:table1}).

From the fact that $\tilde{\gamma}_e(\omega)$ behaves like in the
Ohmic case for small frequencies $\omega \ll \omega_e$, it is also
interesting to consider the leading behavior of $K_e$ for $\omega_e
\to \infty$; from eq. (\ref{eq:exponential_K_q}) we can easily get
$\lim_{\omega_e \to \infty}\,K_e = \frac{\hbar\, \gamma_o}{2\,\pi}
\ne 0$, which is also different from $\lim_{\omega_d \to
\infty}\,K_d$ in Sec. \ref{sec:drude_model}. This confirms that
these limiting values cannot reveal the Ohmic counterpart $K_{0}$.

\begin{table}[htb]
\caption{$K_e/E_g$ for various pairs $(\gamma_o, \omega_e)$, where
$E_g = \frac{\hbar}{2}$ (i.e., $\omega_0 = 1$); $\lim_{\omega_e \to
\infty}\,K_e/E_g = \gamma_o/\pi$. \label{tab:table1}}
\begin{center}
\begin{tabular}{|c||cccc|}
    \hline
    $\omega_e$ & $\gamma_o = 0.5$ & $\gamma_o = 1$ & $\gamma_o = 2$ & $\gamma_o = 5$\\
    \hline
    0.5 & 0.04225 & 0.08186 & 0.15604 & 0.34038\\
    1 & 0.06130 & 0.11838 & 0.22117 & 0.47348\\
    5 & 0.10600 & 0.20348 & 0.37899 & 0.81614\\
    10 & 0.12131 & 0.23326 & 0.43819 & 0.96224\\
    50 & 0.14414 & 0.28018 & 0.54302 & 1.25567\\
    80 & 0.14789 & 0.28896 & 0.56377 & 1.32020\\
    \hline
    \hline
    $\infty$ & 0.15915 & 0.31831 & 0.63662 & 1.59155\\
    \hline
\end{tabular}
\end{center}
\end{table}

\subsection{Extended Ohmic models $(p)$}\label{sec:extended_ohmic_model}
Let us consider damping models with $J_p(\omega) =
M\,\gamma_o\,\omega\,(\omega/\gamma_o)^p$ being polynomially
divergent with $\omega$. Clearly, the case of $p = 0$ is Ohmic.
First, we have $J_1(\omega) = M\,\omega^2$. By using the
relationship $\int_0^{\infty} d y\, e^{i k y} = \pi\,\delta(k) + i\,
P(1/k)$, we can easily obtain $\gamma_1(t) = -\frac{2}{\pi}\,
P\frac{1}{t^2}$, which leads to no well-defined
$\tilde{\gamma}_1(\omega)$. This ($p = 1$) is, therefore, physically
not acceptable. It is not difficult to show that the cases of $p$
being odd are not acceptable.

Next, we consider the case of $p = 2$. It can be shown that
$\gamma_2(t) = -\frac{2}{\gamma_o}\,\delta''(t)$ and
$(\tilde{\gamma}_+)_2(\omega) = \frac{\omega^2}{\gamma_o} -
i\,\frac{2\,\delta(0)}{\gamma_o}\,\omega$. By using this for eq.
(\ref{eq:second_law_inequality_conti}), we can obtain
\begin{equation}\label{eq:higher_ohm1}
    K_2\; =\; \frac{\hbar}{2\,\pi}\; \mbox{Im} \int_0^{\infty}
    d\omega\, \frac{2\, \omega^2\, \{-\omega + i\,\delta(0)\}}{\omega^3 -
    i\,\alpha\,\omega^2 + i\,\beta}\,,
\end{equation}
where $\alpha = \gamma_o + 2\,\delta(0)$ and $\beta =
\omega_0^2\,\gamma_o$. The integral in (\ref{eq:higher_ohm1})
diverges logarithmically. This divergence is, obviously, from the
fact that both $(E_s)_2(0)$ and $F_2(0)$ diverge logarithmically,
however, differently from the Ohmic case, $(E_s)_2(0) \ne F_2(0)$.
In fact, we find that $K_2 = -\frac{\hbar}{\pi}\,\{\delta(0) +
\gamma_o\} \times \infty < 0$, which clearly means that the excess
energy, $(E_s)_2(0)$, is greater than the minimum work ({\em or} the
work in a reversible process), $F_2(0)$, required to couple a system
to a bath. This violation of the second law in the reversible
process may be understood to emerge from a large amount of the
energy offer by the bath with $J_2(\omega)$ diverging with $\omega$.
The infinite value of $K_2$ suggests, however, that this model would
be strictly unrealistic.
%

\subsection{Extended Drude models $(d,n)$}\label{sec:extended_drude_model}
We now consider a more general class of the spectral density than
$J_d(\omega)$, which is
\begin{equation}\label{eq:general_form_spectral_density1}
    \textstyle J_{d,n}(\omega)\; =\;
    \left(\frac{\omega}{\omega_d}\right)^n\, J_d(\omega)\; =\;
    M \gamma_o\,\frac{\omega^{n+1}}{\omega_d^{n-2}\,(\omega^2\, +\,
    \omega_d^2)}\,.
\end{equation}
Let us begin with $n$ being odd. First, $n = 1$. We then have
$J_{d,1}(\omega) = M \gamma_o\,\omega_d\,\omega^2/(\omega^2 +
\omega_d^2)$, which converges to a non-zero constant $M
\gamma_o\,\omega_d$ for large frequencies. Substituting this into
eq. (\ref{eq:damping_kernel2}), we can obtain, after some
calculation (see Appendix~\ref{sec:appendix3} for details),
\begin{eqnarray}\label{eq:gamma_n_1_extended_drude}
    \gamma_{d,1}(t) &=& \frac{\gamma_o\,\omega_d}{\pi}\,
    \left[\,\{\mbox{Ei}(\omega_d t) + E_1(\omega_d t)\}\,\sinh(\omega_d t)\; -\right.\n\\
    && \left.\{\mbox{Ei}(\omega_d t) - E_1(\omega_d t)\}\,\cosh(\omega_d
    t)\,\right]\,.
\end{eqnarray}
Applying the Laplace transform \cite{ROB66} to this, we can get
\begin{equation}\label{eq:laplace_n_1_extended_drude}
    \textstyle (\tilde{\gamma}_+)_{d,1}(\omega)\; =\;
    \frac{\gamma_o\,\omega_d\,\omega}{\omega^2 + \omega_d^2}\, +\,
    i\,\frac{2}{\pi}\,\frac{\gamma_o\,\omega_d\,\omega}{\omega^2 +
    \omega_d^2}\, \ln\left(\frac{\omega}{\omega_d}\right)\,.
\end{equation}
Using eq. (\ref{eq:branch1_1}) with
(\ref{eq:laplace_n_1_extended_drude}), we can arrive at the
expression in (\ref{eq:second_law_inequality_conti})
\begin{equation}\label{eq:k_q_n_1_extended_drude}
    K_{d,1}\; =\; \frac{\hbar \gamma_o}{2\,\pi}\;
    \mbox{Im}\int_0^{\infty} d \lambda\,\frac{g_1(\lambda)}{g_2(\lambda)}\,,
\end{equation}
where
\begin{eqnarray}\label{eq:numerator1_n_1_extended_drude}
    g_1(\lambda) &=& \textstyle \lambda^2\,\left[\,\frac{2\,\lambda_d}{\pi}\,\{(\lambda_d^2 -
    \lambda^2)\, \ln\left(\frac{\lambda}{\lambda_d}\right)\,
    +\, \lambda^2\, +\, \lambda_d^2\}\; +\right.\n\\
    && \left.i\,\lambda_d\,(\lambda^2 - \lambda_d^2)\,\right]\,,\n\\
    g_2(\lambda) &=& \textstyle (\lambda^2 + \lambda_d^2)\,\left\{(\lambda^2 -
    \lambda_0^2)\,(\lambda^2 + \lambda_d^2)\; -\right.\n\\
    && \textstyle \left.\frac{2\,\lambda_d\,\lambda^2}{\pi}\,
    \ln\left(\frac{\lambda}{\lambda_d}\right)\, +\, i\,\lambda_d\,\lambda^2\right\}\,.
\end{eqnarray}
Here, we introduced a dimensionless variable $\lambda =
\omega/\gamma_o$ with $\lambda_0 = \omega_0/\gamma_o$ and $\lambda_d
= \omega_d/\gamma_o$. We numerically evaluate $K_{d,1}$ for various
pairs $(\lambda_0, \lambda_d)$ to show that $K_{d,1} > 0$ (see Table
\ref{tab:table2}). It is also noted that in the limit $\lambda_d$
(or $\omega_d$) $\to \infty$, the spectral density $J_{d,1}(\omega)$
with $\gamma_o = \omega_d$ reduces to $J_{1}(\omega)$ in Sec.
\ref{sec:extended_ohmic_model} for small frequencies. As was
discussed, however, this case $(d,1)$ is a well-defined damping
model whereas the model with $J_1(\omega)$ is not.

\begin{table}[htb]
\caption{$K_{d}\; \pi/\gamma_o E_g$ from
(\ref{eq:drude_k_q__old_parameter_appendix}) versus $K_{d,1}\;
\pi/\gamma_o E_g$ from (\ref{eq:k_q_n_1_extended_drude}) for various
pairs $(\omega_0, \omega_d)$ with $\gamma_o = 1$, where $E_g =
\frac{\hbar\,\omega_0}{2}$; $(d,0)$ denotes the Drude model.
\label{tab:table2}}
\begin{center}
\begin{tabular}{|c|c|c||c|c|c|}
    \hline
    $(\omega_0, \omega_d)$ & $(d, 0)$ & $(d, 1)$ &
    $(\omega_0, \omega_d)$ & $(d, 0)$ & $(d, 1)$\\
    \hline\hline
    $(0.5, 0.5)$ & $0.84275$ & $0.50184$ &
    $(1, 0.5)$ & $1.21124$ & $0.25468$\\
    \hline
    $(0.5, 1)$ & $0.61942$ & $0.45203$ &
    $(1, 1)$ & $0.45318$ & $0.26521$\\
    \hline
    $(0.5, 5)$ & $1.29483$ & $0.34454$ &
    $(1, 5)$ & $0.61427$ & $0.17800$\\
    \hline
    $(0.5, 10)$ & $1.52266$ & $0.24507$ &
    $(1, 10)$ & $0.73767$ & $0.05454$\\
    \hline\hline
    $(5, 0.5)$ & $0.53089$ & $0.02832$ &
    $(10, 0.5)$ & $0.28968$ & $0.00936$\\
    \hline
    $(5, 1)$ & $0.47010$ & $0.04106$ &
    $(10, 1)$ & $0.27637$ & $0.01345$\\
    \hline
    $(5, 5)$ & $0.09790$ & $0.05476$ &
    $(10, 5)$ & $0.13428$ & $0.02346$\\
    \hline
    $(5, 10)$ & $0.09537$ & $0.04976$ &
    $(10, 10)$ & $0.04792$ & $0.02347$\\
    \hline
\end{tabular}
\end{center}
\end{table}

Let $n = 3$ next. We have $J(\omega)_{d,3} = M
\gamma_o\,\omega_d^{-1}\,\omega^4/(\omega^2 + \omega_d^2)$. After a
straightforward calculation, we will obtain
\begin{equation}\label{eq:gamma_n_3_extended_drude}
    \gamma_{d,3}(t)\; =\; -\gamma_{d,1}(t)\, -\,
    \frac{2\,\gamma_o}{\pi\,\omega_d}\, \frac{P}{t^2}\,,
\end{equation}
which indicates that this case is physically not acceptable.
Similarly, we can also show that all cases for $n > 3$ being odd are
not acceptable.

Now, let $n$ be even. We can then find that
\begin{equation}\label{eq:damping_kernel_general_drude1}
    \gamma_{d,2 m}(t)\; =\; (-1)^{m} \left\{\gamma_d(t)\; -\;
    \sum_{j=1}^m
    \frac{\gamma_{0}^{\{2(j-1)\}}(t)}{\omega_d^{2(j-1)}}\right\}\,,
\end{equation}
where $m = 1, 2, \cdots$, and $\gamma_{0}^{\{2(j-1)\}}(t)$ represent
$2(j-1)$-time derivatives of $\gamma_{0}(t)$. We begin with a simple
case $(m = 1)$ with $J_{d,2}(\omega) = M
\gamma_o\,\omega^3/(\omega^2 + \omega_d^2)$. This case is
particularly interesting because $J_{d,2}(\omega)$ diverges for
large frequencies, however, more slowly than $J_0(\omega)$ for the
Ohmic case, whereas all $J_{d,2 m}(\omega)$ for $m > 1$ diverge
faster than $J_0(\omega)$; $J_{d,2}(\omega)$ may be said to be of
{\em weak} divergence. Due to the fact that $K_{d,0}
> K_{d,1} > 0$ seen from Table \ref{tab:table2}, we would like to pose a question if
we will here obtain $K_{d,1} > K_{d,2}
> 0 = K_0$ or $K_{d,1} > 0 \geq K_{d,2}$. In fact, we
have an interesting relation $(\tilde{\gamma}_+)_{d,2}(\omega) =
-\tilde{\gamma}_d(\omega) + \tilde{\gamma}_{0}(\omega)$ from eq.
(\ref{eq:damping_kernel_general_drude1}), and so
$(R_+')_{d,2}(\omega) = -(R_+')_d(\omega)$ for eq.
(\ref{eq:second_law_inequality_conti}). Introducing the parameters
$({\mathbf w}_0, \Omega, \gamma)$ defined as the relations
\begin{equation}\label{eq:parameter_change_extended_drude}
    \textstyle \omega_d\,\omega_0^2 := \Omega\,{\mathbf w}_0^2\,;\;
    \omega_d\, + \gamma_o := \Omega\, +\, \gamma\,;\;
    \omega_0^2 := \Omega\,\gamma\, +\, {\mathbf w}_0^2
\end{equation}
(note that these differ from the relations in
(\ref{eq:parameter_change0})), we can easily obtain
\begin{eqnarray}
    (\tilde{G}_+)_{d,2}(\omega) &=& \frac{\omega^3\, +\, i\,(\gamma_o + \omega_d)\,
    \omega^2\, -\, \omega_0^2\, \omega\, -\, i\,\omega_0^2\,\omega_d}{\omega\, +\, i\,\omega_d}\n\\
    &=& \frac{(\omega\, +\, i \Omega)\,(\omega\, +\, i z_1)\,(\omega\, +\, i z_2)}{\omega\, +\,
    i \Omega\,{\mathbf w}_0^2/(\Omega\,\gamma\, +\, {\mathbf w}_0^2)}\,,\label{eq:susceptibility_extended_drude2}
\end{eqnarray}
where $z_1 = \frac{\gamma}{2} + i\,{\mathbf w}_1$ and $z_2 =
\frac{\gamma}{2} - i\,{\mathbf w}_1$ with ${\mathbf w}_1^2 =
{\mathbf w}_0^2 - (\frac{\gamma}{2})^2$. By using eqs.
(\ref{eq:second_law_inequality_conti}) and
(\ref{eq:susceptibility_extended_drude2}), we arrive at the
expression
\begin{equation}\label{eq:k_q_extended_drude1}
    K_{d,2} \; =\; \frac{\hbar}{2\,\pi}\,\frac{\Omega\,{\mathbf w}_0^2}{(\Omega\,\gamma
    + {\mathbf w}_0^2)\,(\Omega\,\gamma - \Omega^2 - {\mathbf
    w}_0^2)}\; C({\mathbf w}_0, \Omega, \gamma)
\end{equation}
where
\begin{eqnarray}
    C({\mathbf w}_0, \Omega, \gamma) &=&
    \textstyle \gamma\,({\mathbf w}_0^2 - \Omega^2)\, \frac{1}{{\mathbf
    w}_1}\, \arctan \frac{2\,{\mathbf w}_1}{\gamma}\, +\n\\
    && (\Omega^2 + {\mathbf w}_0^2 - \Omega\,\gamma)\, \ln(\Omega\,\gamma +
    {\mathbf w}_0^2)\, -\n\\
    && 2\,(\Omega^2 + {\mathbf w}_0^2)\, \ln {\mathbf w}_0\, +\, 2\,\Omega\,\gamma\, \ln \Omega\,.
\end{eqnarray}
In case that ${\mathbf w}_1$ is complex-valued (${\mathbf w}_0 <
\gamma/2$), i.e., for the overdamped case, this has to be understood
in terms of the relation, $\frac{1}{{\mathbf w}_1} \arctan
\frac{2\,{\mathbf w}_1}{\gamma} = \frac{1}{2\,\bar{{\mathbf {w}}}_1}
\ln\left(\frac{\gamma + 2\,\bar{{\mathbf {w}}}_1}{\gamma -
2\,\bar{{\mathbf {w}}}_1}\right)$, where $\bar{{\mathbf w}}_1^2 =
(\frac{\gamma}{2})^2 - {\mathbf w}_0^2$.

Interestingly enough, we can here observe $K_{d,2} < 0$ (see Fig.
\ref{fig:fig1}), which would allow us to have a violation of the
second law in the reversible process for this damping model. This
negativity may be understood from the comparison, with the aid of
the relation $(R_+')_{d,2}(\omega) = -(R_+')_d(\omega)$, between
\begin{eqnarray}\label{eq:k_q_compare1}
    K_{d,2} &=& \textstyle \frac{\hbar}{2\,\pi}\, \omega_d\,\gamma_o\;
    \times\\
    && \textstyle \mbox{Im} \int_0^{\infty} d\omega\,
    \frac{\omega^2}{(\omega\, +\, i \omega_d)\,(\omega\, +\, i \Omega)\,(\omega\,
    +\, i z_1)\,(\omega\, +\, i z_2)}\n
\end{eqnarray}
with $({\mathbf w}_0, \Omega, \gamma)$  in eq.
(\ref{eq:parameter_change0}) and $\omega_d = \Omega + \gamma$, and
\begin{eqnarray}\label{eq:k_q_compare2}
    K_{d} &=& \textstyle -\frac{\hbar}{2\,\pi}\, \omega_d\,\gamma_o\;
    \times\\
    && \textstyle \mbox{Im} \int_0^{\infty} d\omega\,
    \frac{\omega^2}{(\omega\, +\, i \omega_d)\,(\omega\, +\, i \Omega)\,(\omega\,
    +\, i z_1)\,(\omega\, +\, i z_2)}\, >\, 0\n
\end{eqnarray}
with $({\mathbf w}_0, \Omega, \gamma)$  in
(\ref{eq:parameter_change_extended_drude}) and $\omega_d = {\mathbf
w}_0^2\,\Omega/(\Omega\,\gamma + {\mathbf w}_0^2)$. In fact, we can
also show, by using eq. (\ref{eq:helmholtz_energy1}) with
(\ref{eq:susceptibility_extended_drude2}), that $F_{d,2}(0) >
F_{d}(0)$.

Next, we briefly consider the case of $(d,4)$. We then have
$(\tilde{\gamma}_+)_{d,4}(\omega) = \tilde{\gamma}_{d}(\omega) -
\gamma_o + (\gamma_o\,\omega^2 - 2\,i
\gamma_o\,\delta(0)\,\omega)/\omega_d^2$ from eq.
(\ref{eq:damping_kernel_general_drude1}). After a fairly lengthy
calculation with this, we can eventually obtain an explicit
expression for $K_{d,4} = \frac{\hbar}{2\,\pi} \int_0^{\infty}
d\omega\, f_{d,4}(\omega)$, where $f_{d,4}(\omega) \to
-2\,\{\delta(0) + \omega_d^2/\gamma_o\}/\omega$ for large $\omega$.
From this asymptotic form, we easily see that $K_{d,4} \to -\infty$,
which indicates the violation of the second law. This infinity of
$K_{d,4}$ suggests, however, that this case would be strictly
unrealistic.

From the above results for cases $(d,n)$ including the case $(p =
2)$ in Sec.~\ref{sec:extended_ohmic_model}, we may be able to say
that aside for physically unacceptable damping models, the
divergence (weak or strict) of the spectral density $J(\omega)$ for
large frequencies could lead to the violation of the second law. It
is also interesting here to note that
$\text{Re}\,(\tilde{\gamma}_+)_{d,2}(\omega) =
\gamma_o\,\omega^2/(\omega^2 + \omega_d^2) > 0$ and
$\text{Re}\,(\tilde{\gamma}_+)_{d,4}(\omega) =
\gamma_o\,\omega_d^{-2}\,\omega^4/(\omega^2 + \omega_d^2) > 0$. It
is known \cite{MEI65} that a violation of the positivity for
$\text{Re}\,\tilde{\gamma}_+(\omega)$ is tantamount to a violation
of the second law in the thermodynamic limit (where a coupling
strength between system and bath vanishes). We see here, however,
that this positivity would not be a sufficient condition for the
second law in the quantum regime (with a non-negligible finite
coupling strength between system and bath).\vspace*{.0cm}
%

\section{Conclusions}
In summary, we have extensively studied the second law in the scheme
of quantum Brownian motion. It has been observed that from the
system-bath entanglement, a system oscillator coupled to a bath at
zero temperature has a higher average energy value than the ground
state of an uncoupled harmonic oscillator. For a damping model with
arbitrarily discrete bath modes and damping models with continuous
bath modes with cut-off frequencies, however, this apparent excess
energy has actually been found to be less than the minimum work to
couple a system to a bath. Therefore, the second law holds in the
quantum regime. We also found, on the other hand, that the violation
of the second law may happen for some cut-off frequency-free damping
models, which are, however, physically unrealistic; especially the
case $(d,2)$ in Sec. \ref{sec:extended_drude_model}, with a less
diverging spectral density of bath modes than the Ohmic model being
the prototype for damping, has a finitely negative value of
$K_{d,2}$. The further question about the validity of the quantum
second law for a broader class of quantum systems than the quantum
Brownian motion considered here, particularly non-linear systems
coupled to a bath, clearly remains open.

\section*{Acknowledgments}
One of us (I. K.) is grateful to G.J. Iafrate for some interesting
remarks.

\onecolumn{
\appendix\section{: A detailed derivation of eq. (\ref{eq:second_law_in_general_treatment2})}
\label{sec:appendix1}
By substituting eq. (\ref{eq:susceptibility2}) with
$\tilde{\chi}(\omega) = -1/M G(\omega)$ and eq.
(\ref{eq:gamma_tilde1}) into $K$ in
(\ref{eq:second_law_in_general_treatment1}), we immediately have
\begin{equation}\label{eq:appendix_second_law_in_general_treatment1}
    K\; =\; \frac{-i \hbar}{4 \pi M}\, \oint d\omega\;
    \omega^2\; \frac{{\displaystyle \prod_{j=1}^N\,(\omega^2 - \omega_j^2)}}{{\displaystyle \prod_{k=0}^N\,(\omega^2 -
    \bar{\omega}_k^2)}}\;\, \sum_{l=1}^N\, \frac{c_l^2}{2\,m_l\,\omega_l^2}\;
    \left\{\frac{1}{(\omega_l + \omega)^2}\, +\, \frac{1}{(\omega_l -
    \omega)^2}\right\}\,,
\end{equation}
where the integration path over $\omega$ is a loop around the {\em
positive real axis} in the complex $\omega$-plane, consisting of the
two branches, $(\infty + i \epsilon,\,i \epsilon)$ and $(-i
\epsilon,\,\infty - i \epsilon)$. By using the residues at all poles
$\omega = \bar{\omega}_k$ of the integrand, we can evaluate the
contour integration. In doing so, we do not have any residue at
$\omega = \bar{\omega}_{k'}$ when $\omega_l = \bar{\omega}_{k'}$.
Accordingly, we can finally obtain
\begin{equation}\label{eq:appendix_second_law_in_general_treatment2}
    K\; =\; \frac{\hbar}{4 M}\, \sum_{k=0}^N\,
    \mathcal{A}_k\;\;;\;\;
    \mathcal{A}_k\; =\; \mathcal{A}_k^{(1)} \cdot \mathcal{A}_k^{(2)}\,,
\end{equation}
where
\begin{equation}\label{eq:appendix_second_law_in_general_treatment3}
    \mathcal{A}_k^{(1)}\; =\; \bar{\omega}_k\;
    \frac{\displaystyle \prod_{j=1}^N\, (\bar{\omega}_k^2 - \omega_j^2)}{\displaystyle
    \prod_{k'=0 \atop (\ne k)}^N\, (\bar{\omega}_k^2 -
    \bar{\omega}_{k'}^2)}\;\;;\;\;
    \mathcal{A}_k^{(2)}\; =\; \sum_{l=1}^N\, \frac{c_l^2}{2\,m_l\,\omega_l^2}\;\,
    P\left\{\frac{1}{(\omega_l + \bar{\omega}_k)^2}\, +\, \frac{1}{(\omega_l -
    \bar{\omega}_k)^2}\right\}\,.
\end{equation}
Here, $\mathcal{A}_k^{(2)}$ is, obviously, positive-valued.
Therefore, the non-negativeness of $\mathcal{A}_k$ can be completely
determined by the factor $\mathcal{A}_k^{(1)}$. Keeping in mind the
frequency relationship in (\ref{eq:frequency_relation3}), we first
consider $\mathcal{A}_N^{(1)}$. This is clearly non-negative. Next,
for $k=N-1$, we have
\begin{equation}\label{eq:appendix_second_law_in_general_treatment4}
    \mathcal{A}_{N-1}^{(1)}\; =\; \bar{\omega}_{N-1}\;
    \frac{\displaystyle \prod_{j=1}^{N-1}\, (\bar{\omega}_{N-1}^2 - \omega_j^2)}{\displaystyle
    \prod_{k=0}^{N-2}\, (\bar{\omega}_{N-1}^2 -
    \bar{\omega}_{k}^2)}\; \times\; \frac{\bar{\omega}_{N-1}^2 -
    \omega_N^2}{\bar{\omega}_{N-1}^2 - \bar{\omega}_N^2}\,.
\end{equation}
The first factor on the right hand side is non-negative, and so is
the second factor whose numerator and denominator are
negative-valued, respectively. Therefore, we get $\mathcal{A}_{N-1}
\geq 0$. Along the same line, we can straightforwardly show that
each summand $\mathcal{A}_k$ with $k = N-2, N-3, \cdots, 0$ is
non-negative, which will yield $K \geq 0$.

\section{: No violation of the second law in the Drude model}\label{sec:appendix2}
In the Drude model, we can even evaluate the system energy
$(E_s)_d(0)$ and the free energy $F_d(0)$ explicitly and in closed
form. From eq. (\ref{eq:susceptibility_drude1}) we see that for
${\mathbf w}_0 > \gamma/2$ ({\em underdamped} case), $z_1$ and $z_2$
are conjugate complex numbers to each other, while for ${\mathbf
w}_0 \leq \gamma/2$ ({\em overdamped} case), both $z_1$ and $z_2$
are real-valued. Therefore, $\mbox{Im}\,
\{(R_+')_d(\omega)/(G_+)_d(\omega)\}$ in
(\ref{eq:second_law_inequality_conti}), and thus the explicit
expressions of $K_d$, for both cases would differ from each other in
parameters $({\mathbf w}_0, \Omega, \gamma)$. By using eq.
(\ref{eq:susceptibility_drude1}) we can obtain
\begin{eqnarray}
    \int_0^{\infty} d\omega\; \text{Im} \chi_d(\omega) &=&
    \left\{ \begin{array}{ll}
                \displaystyle \frac{1}{M} \frac{({\mathbf w}_0^2 + \Omega^2 -
                \gamma^2/2)\, \arccos (\gamma/2 {\mathbf w}_0)\, -\, \gamma\,{\mathbf w}_1\,
                \ln (\Omega/{\mathbf w}_0)}{{\mathbf w}_1\, ({\mathbf w}_0^2\, -\, \Omega\,\gamma\,
                +\, \Omega^2)}&\hspace*{.3cm} \text{for}\hspace*{.3cm} {\mathbf w}_0 > \gamma/2\\\label{eq:susceptibility_int1}\\
                \displaystyle \frac{1}{M} \frac{(\gamma^2/2 - {\mathbf w}_0^2 -
                \Omega^2)/2 \cdot \ln \left(\frac{\gamma/2\, -\, \bar{{\mathbf w}}_1}{\gamma/2\,
                +\, \bar{{\mathbf w}}_1}\right)\, -\,
                \gamma\,\bar{{\mathbf w}}_1\, \ln (\Omega/{\mathbf w}_0)}{\bar{{\mathbf w}}_1\,
                ({\mathbf w}_0^2\, -\, \Omega\,\gamma\, +\,
                \Omega^2)}&\hspace*{.3cm} \text{for}\hspace*{.3cm} {\mathbf w}_0 \leq \gamma/2
            \end{array}\right.\\[.3cm]\n\\
    \int_0^{\infty} d\omega\; \omega^2\; \text{Im} \chi_d(\omega) &=&
    \left\{ \begin{array}{ll}
                \displaystyle \frac{1}{M} \frac{({\mathbf w}_0^4 + {\mathbf w}_0^2\,\Omega^2 -
                \Omega^2\,\gamma^2/2)\, \arccos (\gamma/2 {\mathbf w}_0)\,
                +\, \Omega^2\,\gamma\,{\mathbf w}_1\,
                \ln (\Omega/{\mathbf w}_0)}{{\mathbf w}_1\, ({\mathbf w}_0^2\, -\, \Omega\,\gamma\,
                +\, \Omega^2)}&\hspace*{.3cm} \text{for}\hspace*{.3cm} {\mathbf w}_0 > \gamma/2\\\label{eq:susceptibility_int11}\\
                \displaystyle \frac{1}{M} \frac{(\Omega^2\,\gamma^2 - 2\,{\mathbf w}_0^4 -
                2\,{\mathbf w}_0^2\,\Omega^2)/4 \cdot \ln \left(\frac{\gamma/2\, -\,
                \bar{{\mathbf w}}_1}{\gamma/2\, +\, \bar{{\mathbf w}}_1}\right)\, +\,
                \Omega^2\,\gamma\,\bar{{\mathbf w}}_1\,
                \ln (\Omega/{\mathbf w}_0)}{\bar{{\mathbf w}}_1\, ({\mathbf w}_0^2\, -\, \Omega\,\gamma\,
                +\, \Omega^2)}&\hspace*{.3cm} \text{for}\hspace*{.3cm} {\mathbf w}_0 \leq \gamma/2
            \end{array}\right.
\end{eqnarray}
where ${\mathbf w}_1 = \sqrt{{\mathbf w}_0^2 - (\gamma/2)^2}$\, and
$\bar{{\mathbf w}}_1 = -i\,{\mathbf w}_1$.

From eqs. (\ref{eq:excess_energy1}), (\ref{eq:susceptibility_int1}),
and (\ref{eq:susceptibility_int11}), we have an exact expression
\begin{equation}\label{eq:energy_drude1}
    (E_s)_d(0)\; =\; \frac{\hbar}{2 \pi}\, \left\{A({\mathbf w}_0, \Omega, \gamma)\,
    +\, B({\mathbf w}_0, \Omega, \gamma)\right\}\,,
\end{equation}
where
\begin{equation}\label{eq:energy_drude2}
    A({\mathbf w}_0, \Omega, \gamma)\; =\;
    \left\{ \begin{array}{l}
                \displaystyle \frac{({\mathbf w}_0^2 + \Omega^2)\, (2\,\Omega\,{\mathbf w}_1^2\, +\,
                {\mathbf w}_0^2\,\gamma)\, -\,
                \Omega^2\,\gamma^3/2}{{\mathbf w}_1\, (\Omega\, +\, \gamma)\, ({\mathbf w}_0^2\, -
                \Omega\,\gamma\, +\,
                \Omega^2)}\, \arccos(\gamma/2 {\mathbf w}_0)\\\\
                \displaystyle \frac{({\mathbf w}_0^2 + \Omega^2)\, (\Omega\,\gamma^2/4\, -\,
                {\mathbf w}_0^3\, -\, {\mathbf w}_0^2\,\gamma/2)\, -\,
                \Omega\,\gamma^2/2 \cdot ({\mathbf w}_0^2\, -\,
                \Omega\,\gamma/2)}{\bar{{\mathbf w}}_1\,
                (\Omega\, +\, \gamma)\, ({\mathbf w}_0^2\, -\, \Omega\,\gamma\,
                +\, \Omega^2)}\, \ln\left(\frac{\gamma/2\, -\, \bar{{\mathbf w}}_1}{\gamma/2\,
                +\, \bar{{\mathbf w}}_1}\right)
            \end{array}\right.
\end{equation}
for ${\mathbf w}_0 > \gamma/2$ and ${\mathbf w}_0 \leq \gamma/2$,
respectively, and
\begin{equation}
    B({\mathbf w}_0, \Omega, \gamma)\; =\; \frac{\Omega\,\gamma\, (\Omega^2\, +\,
    \Omega\,\gamma\, -\, {\mathbf w}_0^2)}{(\Omega\, +\, \gamma)\, ({\mathbf w}_0^2\, -\,
    \Omega\,\gamma\, +\, \Omega^2)}\, \ln (\Omega/{\mathbf w}_0)\,.\label{eq:energy_drude3}
\end{equation}
Similarly, the free energy $F_d(0)$ in eq.
(\ref{eq:helmholtz_energy1}) can be exactly evaluated as
\begin{equation}\label{eq:helmholtz_energy2_appendix}
    F_d(0)\; =\;
    \left\{ \begin{array}{ll}
                \displaystyle \frac{\hbar}{2 \pi}\, \left\{(\Omega +
                \gamma)\, \ln\left(\frac{\Omega + \gamma}{\Omega}\right)\,
                +\, \gamma\, \ln\left(\frac{\Omega}{{\mathbf w}_0}\right)\, +\, 2\,{\mathbf w}_1\,
                \arccos \left(\frac{\gamma}{2\,{\mathbf w}_0}\right)\right\}&\hspace*{.3cm} \text{for}\hspace*{.3cm}
                {\mathbf w}_0 > \gamma/2\\\\
                \displaystyle \frac{\hbar}{2 \pi}\, \left\{(\Omega + \gamma)\,
                \ln\left(\frac{\Omega + \gamma}{\Omega}\right)\,
                +\, \gamma\, \ln\left(\frac{\Omega}{{\mathbf w}_0}\right)\,
                +\, \bar{{\mathbf w}}_1\,
                \ln \left(\frac{\gamma/2\, -\, \bar{{\mathbf w}}_1}{\gamma/2\,
                +\, \bar{{\mathbf w}}_1}\right)\right\}&\hspace*{.3cm} \text{for}\hspace*{.3cm} {\mathbf w}_0 \leq
                \gamma/2\,.
            \end{array}\right.
\end{equation}
Clearly, $(E_s)_d(0)$ and $F_d(0)$ for the underdamped case are
identical to eqs. (10) and (14) in \cite{FOR06}, respectively. These
expressions can also be applied for the overdamped case (${\mathbf
w}_1 \not\in {\mathbb R}$) with the aid of the complex-valued
expression, $\arccos(y) = \frac{\pi}{2} + i\,\ln(i\,y + \sqrt{1 -
y^2})$ and actually equivalent to those for the overdamped case
derived here. Then, we can obtain $E_s(0) < F(0)$ for both cases
(see Fig. 2 in \cite{FOR06}).

In the limit $\Omega \to \infty$ (equivalently, $\omega_d \to
\infty$), we get from eqs.
(\ref{eq:energy_drude1})-(\ref{eq:helmholtz_energy2_appendix})
\begin{equation}\label{eq:k_q_drude_omega_to_infinity1}
    K_d\; \approx\; \frac{\hbar\,\Omega}{2 \pi}\,
    \ln\left(\frac{\Omega + \gamma}{\Omega}\right)\; \approx\;
    \frac{\hbar\,\gamma}{2 \pi}\,
    \left(1 - \frac{\gamma}{2 \Omega}\right)\, +\,
    \mathcal{O}\left(\frac{1}{\Omega^2}\right)
\end{equation}
which reduces to
\begin{equation}\label{eq:k_q_drude_omega_to_infinity2}
    \frac{\gamma}{\pi {\mathbf w}_0} E_g\; =\;
    \frac{\hbar\,\gamma}{2 \pi}\,\frac{1}{\sqrt{1 + \gamma/\Omega}}\;
    \approx\; \frac{\hbar\,\gamma}{2 \pi}\,\left(1 -
    \frac{\gamma}{2 \Omega}\right)\, +\,
    \mathcal{O}\left(\frac{1}{\Omega^2}\right)
\end{equation}
({\em cf}. eq.~(15) in \cite{FOR06}). Lastly, we explicitly give an
explicit expression for $K_d$ in original parameters $(\omega_0,
\omega_d, \gamma_o)$,
\begin{equation}\label{eq:drude_k_q__old_parameter_appendix}
    K_d\; =\; \frac{\hbar\,\gamma_o}{2 \pi}\, \int_0^{\infty}
    d\lambda\;
    \frac{2\,\lambda_d^2\,\lambda^5\, -\, (2\,\lambda_0^2\, +\, \lambda_d)\,\lambda_d^2\,\lambda^3}{\{(\lambda^2 + \lambda_d^2)\,(\lambda^2 -
    \lambda_0^2)\, -\, \lambda_d\,\lambda^2\}^2\, +\, (\lambda_d^2\,\lambda)^2}\,,
\end{equation}
where a dimensionless parameter $\lambda = \omega/\gamma_o$ with
$\lambda_0 = \omega_0/\gamma_o$ and $\lambda_d = \omega_d/\gamma_o$.
Here, we see that the non-zero value of $K_d$ for $\lambda_d \to
\infty$ arises from the competition between two terms of the
numerator of the integrand for large $\lambda$. Eq.
(\ref{eq:drude_k_q__old_parameter_appendix}) is also used in Table
\ref{tab:table2} for comparison with $K_{d,1}$ in eq.
(\ref{eq:k_q_n_1_extended_drude}).

\section{: Details for eqs. (\ref{eq:exponential_frequency1}) and (\ref{eq:gamma_n_1_extended_drude})}\label{sec:appendix3}
In derivation of eq. (\ref{eq:exponential_frequency1}), we used the
relationship\cite{ABS74} $E_1(-y \pm i\,0^+) = -\mbox{Ei}(y) \mp
i\,\pi$, where the exponential integrals $E_1(y) = \int_1^{\infty} d
z\, e^{-y\,z}/z$ and $\mbox{Ei}(y) = P \int_{-\infty}^y d z\,
e^z/z$. Also, for eq. (\ref{eq:gamma_n_1_extended_drude}), we
employed, first, \cite{GRA00}
\begin{equation}\label{eq:integral_n_1_1}
    \int_0^{\infty} dy\, \frac{\cos(a\,y)}{y\, +\, b}\; =\;
    -\sin(a\,b)\; \mbox{si}(a\,b)\, -\, \cos(a\,b)\; \mbox{Ci}(a\,b)\,,
\end{equation}
where the sine integral\, $\mbox{si}(y) = -\int_y^{\infty}
dz\,\frac{\sin(z)}{z} = -\frac{\pi}{2} + \mbox{Si}(y)$ with
$\mbox{Si}(y) = \int_0^y dz\,\frac{\sin(z)}{z}$, and the cosine
integral $\mbox{Ci}(y) = -\int_y^{\infty} dz\,\frac{\cos(z)}{z} =
c_e + \ln y + \int_0^y dz\,\frac{\cos(z) - 1}{z}$ with the Euler
constant $c_e = 0.5772156649 \cdots$; secondly, we used the
relations $\mbox{Si}(i y) = \frac{i}{2} \{\mbox{Ei}(y) + E_1(y)\}$
and $\mbox{Ci}(i y) = \frac{1}{2} \{\mbox{Ei}(y) - E_1(y)\} +
\frac{\pi}{2}i$ \cite{ABS74}.
}
%

\begin{figure}[htb]
\centering \hspace*{-5.6cm}\includegraphics{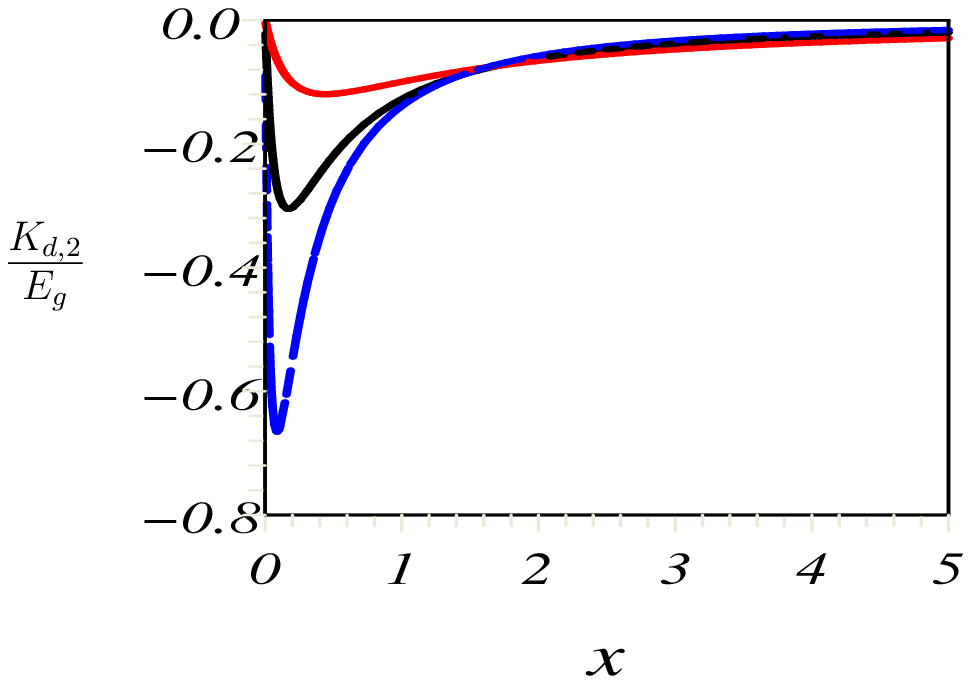}
\caption{\label{fig:fig1}}
\end{figure}

\twocolumn{

Fig.~\ref{fig:fig1}: $K_{d,2}/E_g$ versus $x = \gamma/{\mathbf
w}_0$, where the ground state energy $E_g =
\frac{\hbar}{2}\,\sqrt{\Omega\,\gamma + {\mathbf w}_0^2}$; for
$\Omega = 2\,{\mathbf w}_0$ (dot),\, $5\,{\mathbf w}_0$ (solid),\,
$10\,{\mathbf w}_0$ (dashed) from top to bottom.
}

\begin{thebibliography}{1}
\bibitem{IKI06} Following, e.g., Refs. \cite{HAE05}, \cite{HAE06}, \cite{ING98}, and \cite{LEV88},
we adopt the Laplace transforms for an analysis of the frequency
domain in this paper, instead of the Fourier transforms which have
been employed in a series of papers by Ford and O'Connell, e.g.,
Refs. \cite{FOR06}, \cite{FOR05}, and \cite{FOR85}. Clearly, both
are a priori equally acceptable to give rise to the same desired
physical results in the end.
\end{thebibliography}
\end{document}